\documentclass[final,5p,times,twocolumn]{elsarticle}

\usepackage[utf8]{inputenc}
\usepackage[T1]{fontenc}
\usepackage{float}
\usepackage{verbatim}
\usepackage{apalike}
\usepackage{hyperref}
\usepackage{url}
\usepackage{booktabs}
\usepackage{amsfonts}
\usepackage{nicefrac}
\usepackage{microtype}
\usepackage{lipsum}
\usepackage{graphicx}
\usepackage{tikz}
\usepackage{amssymb}
\usepackage{amsmath}
\usepackage{cleveref}
\usepackage{lineno}
\usepackage{multirow}
\usepackage{cuted}
\usepackage{enumitem}
\usepackage{array}
\usepackage{tabularx}
\usepackage{adjustbox}

\graphicspath{{./}}

\restylefloat{figure}
\floatstyle{plaintop}
\restylefloat{table}

\biboptions{authoryear,round,semicolon}
\AtBeginDocument{\let\cite\citep}

\journal{arXiv preprint}

\newcommand{\titletext}{Reliable Hierarchical Operating System Fingerprinting via Conformal Prediction}

\newcommand{\orcidicon}{\includegraphics[scale=0.015]{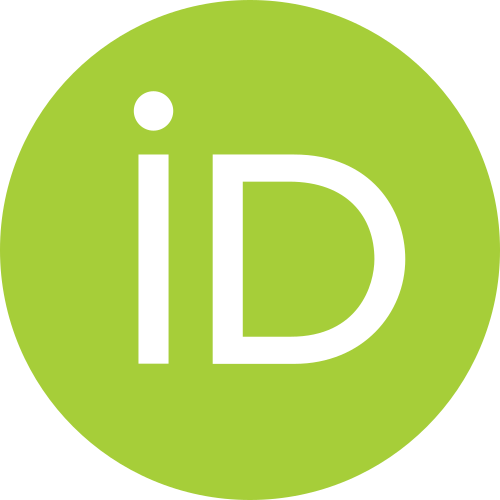}\hspace{2pt}}

\newcommand{\AuthorsFull}{
	\author[1,2]{Rubén Pérez-Jove\corref{cor1} \href{https://orcid.org/0000-0002-7988-945X}{\orcidicon}}
	\ead{ruben.perez.jove@udc.es}

	\author[3]{Osvaldo Simeone \href{https://orcid.org/0000-0001-9898-3209}{\orcidicon}}
	\ead{o.simeone@northeastern.edu}

	\author[1,2]{Alejandro Pazos \href{https://orcid.org/0000-0003-2324-238X}{\orcidicon}}
	\ead{alejandro.pazos@udc.es}

	\author[1,2]{Jose Vázquez-Naya \href{https://orcid.org/0000-0002-6194-5329}{\orcidicon}}
	\ead{jose@udc.es}

	\cortext[cor1]{Corresponding author.}

	\address[1]{RNASA-IMEDIR, Universidade da Coruña, Facultad de Informática, Campus de Elviña, A Coruña 15071, Spain}
	\address[2]{Centro de Investigación CITIC, Universidade da Coruña, Campus de Elviña, A Coruña 15071, Spain}
	\address[3]{Intelligent Networked Systems Institute (INSI), Northeastern University London, One Portsoken Street, London E1 8PH, United Kingdom}
}

\newcommand{\RepoUrl}{https://github.com/rubenpjove/CP-HOSfing}
\newcommand{\RepoLabel}{github.com/rubenpjove/CP-HOSfing}
\newcommand{\RepoLink}{\href{\RepoUrl}{\RepoLabel}}

\newcommand{\abstracttext}{
Operating System (OS) fingerprinting is critical for network security, but conventional methods do not provide formal uncertainty quantification mechanisms. Conformal Prediction (CP) could be directly wrapped around existing methods to obtain prediction sets with guaranteed coverage. However, a direct application of CP would treat OS identification as a flat classification problem, ignoring the natural taxonomic structure of OSs and providing brittle point predictions. This work addresses these limitations by introducing and evaluating two distinct structured CP strategies: level-wise CP (L-CP), which calibrates each hierarchy level independently, and projection-based CP (P-CP), which ensures structural consistency by projecting leaf-level sets upwards. Our results demonstrate that, while both methods satisfy validity guarantees, they expose a fundamental trade-off between level-wise efficiency and structural consistency. L-CP yields tighter prediction sets suitable for human forensic analysis but suffers from taxonomic inconsistencies. Conversely, P-CP guarantees hierarchically consistent, nested sets ideal for automated policy enforcement, albeit at the cost of reduced efficiency at coarser levels.
}

\newcommand{\keywordstext}{
Operating System \sep Fingerprinting \sep Conformal Prediction \sep Hierarchical Classification \sep Uncertainty Quantification \sep Network Security \sep Cybersecurity \sep Deep Learning \sep Machine Learning
}

\newcommand{\creditauthorship}{
\section*{CRediT authorship contribution statement}
\label{sec:credit_authorship}

\begin{itemize}[noitemsep, topsep=2pt]
	\item \textbf{Rubén Pérez-Jove}: Conceptualization; Data curation; Formal analysis; Investigation; Methodology; Software; Validation; Visualization; Writing - original draft; Writing - review \& editing.
	\item \textbf{Osvaldo Simeone}: Conceptualization; Investigation; Methodology; Project administration; Resources; Supervision; Writing - review \& editing.
	\item \textbf{Alejandro Pazos}: Funding acquisition; Project administration; Resources; Supervision.
	\item \textbf{Jose Vázquez-Naya}: Conceptualization; Funding acquisition; Investigation; Methodology; Project administration; Resources; Supervision; Writing - review \& editing.
\end{itemize}
}

\newcommand{\declarationcompetinginterest}{
\section*{Declaration of Competing Interest}
\label{sec:declaration_competing_interest}

The authors declare that they have no known competing financial interests or personal relationships that could have appeared to influence the work reported in this paper.
}

\newcommand{\funding}{
\section*{Funding}
\label{sec:funding}

This work was supported by the grant ED431C 2022/46 - Competitive Reference Groups GRC - funded by \textit{Xunta de Galicia} (Spain). This work was also supported by CITIC, as a center accredited for excellence within the Galician University System and a member of the CIGUS Network, which receives subsidies from the Department of Education, Science, Universities, and Vocational Training of the \textit{Xunta de Galicia}. Additionally, CITIC is co-financed by the EU through the FEDER Galicia 2021–27 operational program (Ref. ED431G 2023/01). This work was also supported by the ``\textit{Formación de Profesorado Universitario}'' (FPU) grant from the Spanish Ministry of Universities to Rubén Pérez Jove (Grant FPU22/04418). This work was supported by the inMOTION programme, INDITEX-UDC Predoctoral Research Stay Grants (2025 call), under the collaboration agreement between Universidade da Coruña (UDC) and INDITEX, S.A. This work was also made possible through the access granted by the Galician Supercomputing Center (CESGA) to its supercomputing infrastructure. The supercomputer FinisTerrae III and its permanent data storage system have been funded by the Spanish Ministry of Science and Innovation, the Galician Government and the European Regional Development Fund (ERDF). Funding for open access charge: Universidade da Coruña/CISUG. The work of O. Simeone was supported by the European Research Council (ERC) under the European Union’s Horizon Europe Programme (grant agreement No. 101198347), by an Open Fellowship of the EPSRC (EP/W024101/1), and by the EPSRC project (EP/X011852/1).
}

\newcommand{\dataavailability}{
\section*{Data availability}
\label{sec:data_availability}

The data and code used in this study is available under the GNU GPL v3 License in the following public GitHub repository: \RepoLink.
}

\hypersetup{
	pdftitle={\titletext},
	pdfauthor={Ruben Perez-Jove; Osvaldo Simeone; Alejandro Pazos; Jose Vazquez-Naya}
}

\newcommand{\BlindedNote}[1]{
}

\begin{document}

	\begin{frontmatter}

		\title{\titletext}

		\AuthorsFull

		\begin{abstract}
			\abstracttext
		\end{abstract}


	\end{frontmatter}

\section{Introduction}
\label{sec:introduction}

\begin{figure*}[t]
	\centering
	\includegraphics[width=\textwidth]{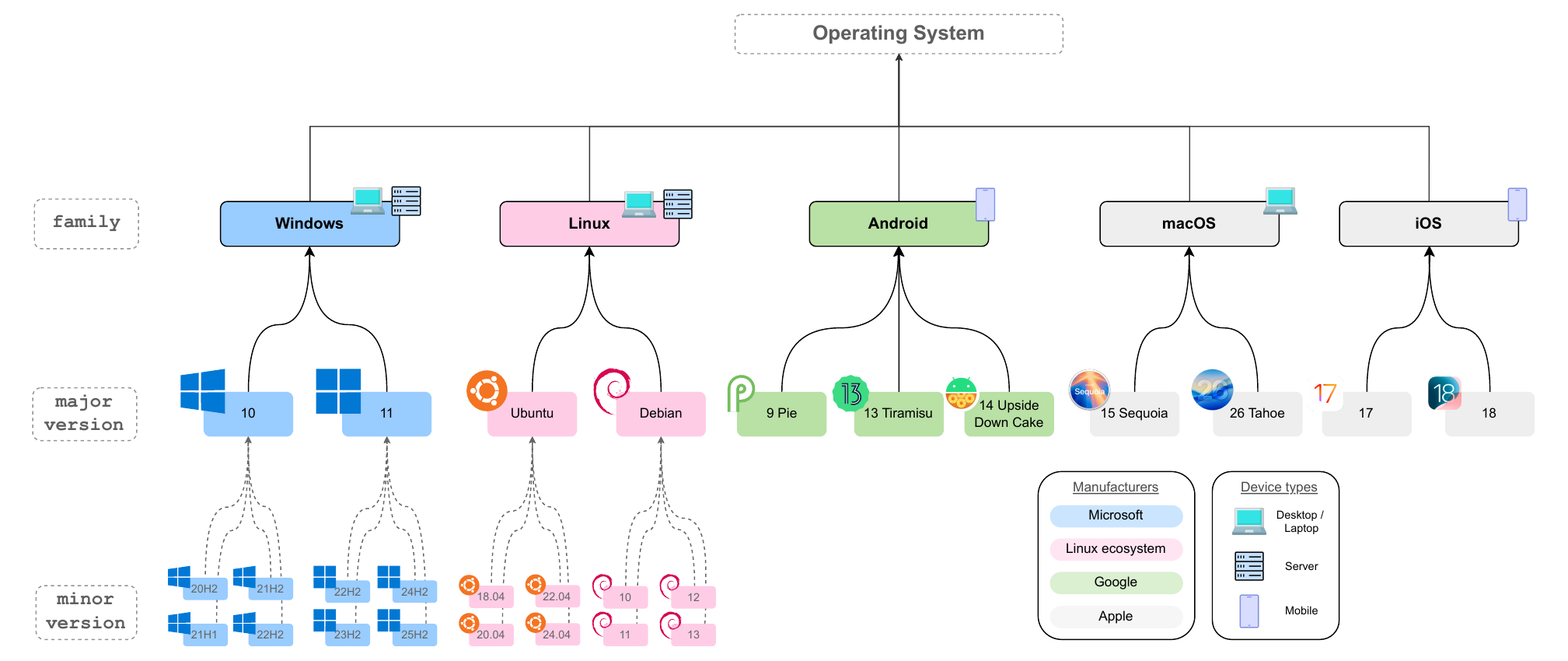}
	\caption{Diagram of an illustrative example of hierarchical label tree in OS fingerprinting. This example adopts three levels---\texttt{OS family} $\rightarrow$ \texttt{major version} $\rightarrow$ \texttt{minor version}---for common current desktop, server and mobile OSs (e.g., \texttt{Windows}, \texttt{Linux} distributions, \texttt{macOS}, \texttt{Android}, \texttt{iOS}). The structure is not canonical but a concrete, non-exhaustive instance aligned with the general $K$-level formulation used in this work. Moreover, the diagram shows that not all OS families expose all three levels; some branches terminate early (e.g., missing minor versions), yielding a typically ragged hierarchy.}
	\label{fig:os_tree}
\end{figure*}

Operating System (OS) fingerprinting is the process of identifying the OS of a remote machine by analysing specific characteristics of its network traffic~\cite{zalewski_p0f_2005}. This capability is fundamental in network monitoring, asset management, and network security. In practice, OS predictions drive two main operational workflows: automated policy enforcement, where systems autonomously grant or deny access based on device type (e.g., BYOD policies), and forensic triage, where human analysts use OS information to route incident tickets to appropriate response teams or prioritise investigations~\cite{lastovicka_passive_2023}.

The process is based on the fact that different OSs and their respective versions implement their network protocol stacks according to standards, but incorporating minor variations. OS fingerprinting methods are divided into active approaches, which probe a device with crafted packets and infer its OS from the responses~\cite{lyon_nmap_2009, veysset_new_2002, li_overview_2020}, and passive approaches, which analyse existing traffic without interacting with the host~\cite{zalewski_p0f_2005, ornaghi_alberto_ettercap_2001, noauthor_networkminer_nodate}.

However, most current solutions rely on point predictions (outputting a single label), which are brittle under the severe class imbalance and open-set nature of network traffic. In high-stakes security contexts like Bring-Your-Own-Device (BYOD) enforcement, a confident but incorrect prediction can lead to security breaches.

A potential solution to this problem is given by Conformal Prediction (CP)~\cite{angelopoulos_conformal_2023}, which wraps around existing classifiers to produce sets of likely candidates (e.g., \texttt{\{Android 10, Android 11\}}) that allow for risk-aware admission policies. However, a direct application of CP to OS fingerprinting would treat it as a flat classification problem, ignoring the inherent hierarchical structure of operating systems (e.g., \texttt{Windows $\to$ Windows 10 $\to$ Windows 10 22H2}) (see \Cref{fig:os_tree} for an illustration). This simplification misses crucial taxonomic information: confusing two versions of the same OS family is operationally less severe than misclassifying the family entirely.

To address these limitations, this work exploits the natural taxonomic organisation of OS labels by introducing and evaluating two distinct structured CP strategies: level-wise CP (L-CP) and projection-based CP (P-CP).

\begin{figure*}[t]
	\centering
	\includegraphics[width=\textwidth]{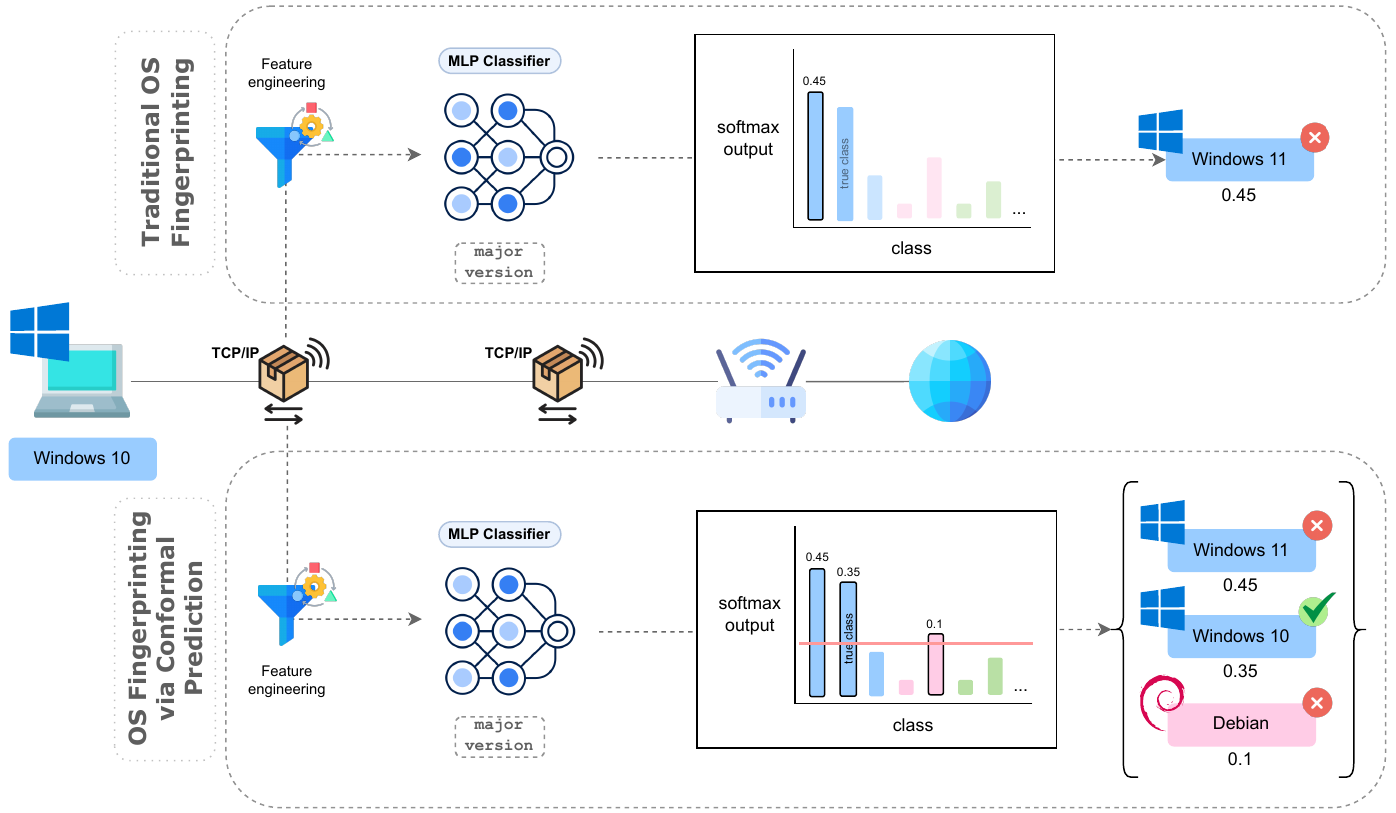}
	\caption{Comparison of traditional OS fingerprinting and OS fingerprinting with Conformal Prediction (CP). The top section illustrates the conventional approach, where a single point prediction (\texttt{Windows 11}) is made based on the highest softmax probability, which may be incorrect. The bottom section demonstrates how CP generates a prediction set containing multiple plausible OS labels, including the true class (\texttt{Windows 10}) even when it is not the highest probability output. This set-valued approach provides calibrated uncertainty quantification, enabling risk-aware decision-making in network security applications.}
	\label{fig:osfing_vs_cp_osfing}
\end{figure*}

The main contributions are as follows:

\begin{itemize}
    \item \textbf{Hierarchical CP Application:} We present the first application of CP to hierarchical OS fingerprinting. We address the specific challenges of taxonomic consistency and multi-level uncertainty in network security by presenting L-CP and P-CP, two structured variants of CP (see \Cref{fig:osfing_vs_cp_osfing}).
    \item \textbf{Comprehensive Evaluation Framework:} We introduce a rigorous comparison of CP strategies using both standard efficiency metrics (coverage, set size, etc.) and a novel structural metric, the \textit{Hierarchical Inconsistency Rate} (HIR), designed to quantify logical inconsistencies in hierarchical predictions.
    \item \textbf{Trade-off Analysis:} We perform an empirical analysis on real-world traffic demonstrating the trade-off between level-wise efficiency (i.e., tightness of prediction sets) and structural consistency (i.e., logical coherence across hierarchy levels) to guide operational deployment. Specifically, L-CP calibrates each hierarchy level independently (see \Cref{fig:l_cp}), and P-CP ensures structural consistency by projecting leaf-level sets upwards (see \Cref{fig:p_cp}).
    \item \textbf{Reproducibility:} We provide a complete, open-source implementation of the proposed framework and the experimental pipeline in a public GitHub repository (\RepoLink) to foster reproducibility and future research.
\end{itemize}

The remainder of this paper is organised as follows. \Cref{sec:background} reviews the evolution of OS fingerprinting and the fundamentals of CP. \Cref{sec:problem_formulation} formally defines the hierarchical classification problem and the CP strategies. \Cref{sec:materials_methods} details the dataset, base models, and evaluation metrics. \Cref{sec:results_discussion} presents the experimental results and discusses the operational implications. Finally, \Cref{sec:conclusion} concludes the paper.

\section{Background}
\label{sec:background}

This section provides the necessary background to contextualise our work. We first review the evolution of OS fingerprinting techniques, from early signature-based approaches to modern machine learning methods, highlighting the challenges posed by encrypted traffic and class imbalance. We then introduce the fundamentals of CP, a framework for uncertainty quantification that forms the theoretical foundation for our hierarchical prediction sets.

\subsection{OS Fingerprinting}
\label{subsec:background_os_fing}

Classic OS fingerprinting methods are based on analyzing differences between the implementations that different OSs make of the TCP/IP protocols. Specifically, they focused on analyzing the headers of such protocols (\cite{postel_internet_1981, postel_transmission_1981}). These traditional approaches, exemplified by tools like Nmap~\cite{lyon_nmap_2009} or p0f~\cite{zalewski_p0f_2005}, exploit parameters such as the initial Time To Live (TTL), TCP window size, Maximum Segment Size (MSS), or the length of the initial Synchronize (SYN) packet~\cite{matousek_towards_2014, lastovicka_passive_2018, lastovicka_passive_2018-1, hagos_machine-learning-based_2021, lastovicka_machine_2018, Perez-Jove-2023-OperatingSystemFingerprinting-conferencePaper}. They typically rely on predefined, expert-crafted signature databases that match protocol‐stack idiosyncrasies to OS labels.

However, Machine Learning (ML) has been increasingly adopted to go beyond static signatures. Some of these works have leveraged application-layer identifiers (e.g., HTTP headers)~\cite{kline_structure_2017, lastovicka_passive_2018} with the aim of achieving better identification. However, the widespread adoption of encryption has obscured these traditional markers. Nevertheless, unencrypted handshake parameters remain accessible, such as TLS Client Hello features~\cite{husak_https_2016, anderson_accurate_2020, fan_identify_2019, lastovicka_using_2020}, flow metadata, and OS-specific domains (e.g.,
\texttt{update.microsoft.com} in Windows or \texttt{connectivitycheck.android.com} in
Android)~\cite{chang_study_2015, lastovicka_passive_2018}. Combinations of TLS, DNS, and flow statistics are now commonly used to enrich the feature space in encryption-dominated networks~\cite{husak_https_2016, anderson_accurate_2020, shamsi_faulds_2021,voronov_determining_2021, lastovicka_passive_2023}.

While classical signature-based methods were sufficient in simpler environments, the exponential increase in OS versions makes them increasingly brittle and difficult to maintain with manual expert knowledge. The shift towards ML and modern features offers a more robust solution, as learning discriminative patterns directly from traffic is easier to adapt and better positioned for future network evolutions~\cite{Perez-Jove-2026-Networktrafficfoundation-journalArticle}.

From the ML perspective, OS fingerprinting is typically cast as a supervised multi-class classification task, with labels defined at different non-standard granularities (e.g., \texttt{family}, \texttt{major}, or \texttt{minor version}). Training relies on labelled datasets obtained in controlled settings or inferred from application-layer identifiers (e.g., Dynamic Host Configuration Protocol (DHCP) logs, Hypertext Transfer Protocol (HTTP) User-Agent), which raises scalability and generalisation issues as new versions appear that are absent from existing corpora.

The label distribution is markedly imbalanced (\cite{li_passive_2023}), where dominant families such as Windows or Android overshadow rarer Linux distributions or legacy systems. This implies re-sampling, cost-sensitive learning, or calibration to maintain reliable performance.

In addition, the heterogeneity of available features, ranging from TCP/IP header parameters to TLS handshake fields and DNS metadata, makes OS fingerprinting a complex learning problem. Models in this field must integrate categorical, numerical, and sequential information, and their performance is typically evaluated using standard metrics such as accuracy, precision, recall, and F1-score.

In deployment, however, the problem is inherently open-set: models encounter unseen families/versions, so methods should be able to abstain, adapt, or detect novelty and remain robust to distributional shift in dynamic networks (where characteristics change over time: OS distribution, used protocols, etc.). Evaluation must therefore go beyond to include reliability metrics such as coverage. This necessity for reliable uncertainty quantification motivates the application of CP.

\subsection{Conformal Prediction}
\label{subsec:background_cp}

Conformal Prediction (CP) provides distribution-free, finite-sample uncertainty quantification layered on top of any underlying learning algorithm (the base predictor). Under the mild assumption of exchangeability of the data, CP yields set-valued predictions with guaranteed marginal coverage at a user-specified level~\cite{angelopoulos_conformal_2023}. Unlike traditional point predictions that output a single label, CP generates prediction sets that may contain multiple plausible labels.

\paragraph*{Mathematical formulation}
Let $(X_i,Y_i)_{i=1}^{n}$ be exchangeable training examples with $X_i\in\mathcal{X}$ and $Y_i\in\mathcal{Y}$, and let $(X_{n+1},Y_{n+1})$ be an exchangeable test pair. In the split (inductive) CP setting, we fit any probabilistic or score-based classifier $f$ on a proper training subset, and we keep aside a calibration set $\mathcal{C}=\{(X_i,Y_i)\}_{i=1}^{m}$. A nonconformity score $s:\mathcal{X}\times\mathcal{Y}\to\mathbb{R}$ measures how atypical a label $y$ is for features $x$ according to $f$ (e.g., $s(x,y)=1-p_\theta(y\mid x)$ for a softmax model, or a ranking-/cumulative-probability score). Compute calibration scores $A_i=s(X_i,Y_i)$ for $(X_i,Y_i)\in\mathcal{C}$, and let

\begin{equation}
	\widehat{q}_{1-\alpha}
	\triangleq
	\left(\cdot\right)\!\left(
	\{A_i\}_{i=1}^{m}\,;\,
	\frac{\left\lceil (m+1)(1-\alpha)\right\rceil}{m+1}
	\right).
	\label{eq:cp-q}
\end{equation}

\noindent where $\left(\cdot\right)$ denotes the right-continuous empirical quantile. Equivalently, $\widehat{q}_{1-\alpha}$ is the $\lceil (m+1)(1-\alpha)\rceil$-th smallest number in the set of $\{A_i\}_{i=1}^m$.

For a new input $x$, the conformal prediction set is

\begin{equation}
	\Gamma_\alpha(x)\;=\;\big\{\, y\in\mathcal{Y}:\ s(x,y)\le \widehat{q}_{1-\alpha}\,\big\}.
	\label{eq:cp-set}
\end{equation}

Then, without assumptions on the data distribution or the correctness of the model $f$, CP guarantees the marginal coverage condition

\begin{equation}
	\mathbb{P}\!\big\{\,Y_{n+1}\in \Gamma_\alpha(X_{n+1})\,\big\}\;\ge\;1-\alpha,
	\label{eq:cp-validity}
\end{equation}

\noindent where the probability is over all data and the test point, assuming exchangeability~\cite{angelopoulos_conformal_2023}.

\section{Problem Formulation}
\label{sec:problem_formulation}

Having established the background, we now formalise the hierarchical OS fingerprinting problem and present two distinct strategies for applying CP to this setting. This section first characterises the hierarchical structure of OS labels and its implications for classification, then introduces the mathematical framework for both level-wise CP (L-CP) and projection-based CP (P-CP), highlighting their respective guarantees and limitations.

\subsection{Hierarchical OS Fingerprinting}
\label{subsec:hierarchical_os_fing}

A key characteristic of OS fingerprinting is its inherently hierarchical structure. OSs can be organised taxonomically into a rooted tree with $K$ levels (e.g., \texttt{OS family} at level one, \texttt{major version} at level two, and \texttt{minor version} at level three or the leaves), which allows the task to be cast as hierarchical classification (see \Cref{fig:os_tree}). This framing acknowledges that not all errors are equally severe: confusing \texttt{Windows~10} with \texttt{Windows~11} is less detrimental than misclassifying \texttt{Windows} as \texttt{Android}. Hierarchical methods, or multi-task learning formulations, that jointly predict the labels at different granularities, are therefore well-suited to exploit shared representations and improve generalisation to under-represented classes.

In this context, the number of levels ($K$ levels of the tree) is problem- and dataset-dependent, and is generally determined by:

\begin{itemize}[label={},leftmargin=1em]
\item (i) the semantic granularity available in the labels of each dataset (e.g., whether the labelling techniques or dataset information distinguishes only family and version, or also fine-grained minor builds), and
\item (ii) the way labels are encoded (some intermediate levels are not standardised across OS families and may be collapsed or represented differently).
\end{itemize}

In practice, hierarchies are often ragged: certain families may lack information at finer levels (e.g., no minor-build annotations for specific macOS releases), yielding variable depth across branches. Consequently, prior work may legitimately adopt two levels (\texttt{family/version}), three levels (\texttt{family/major/minor}), or more. \Cref{fig:os_tree} represents an illustrative simplified example of one realistic scenario, but not the general or unique case.

Viewing OS fingerprinting as a hierarchical classification problem offers several advantages. First, it mirrors the natural taxonomic organisation of OSs, enabling models to exploit dependencies between classes. Second, it provides flexible granularity: in some use cases, identifying only the OS family may be sufficient (e.g., for BYOD policy enforcement), whereas vulnerability detection often requires distinguishing between major or minor versions. Third, hierarchical classification enables more informative evaluation, since errors between closely related classes (e.g., \texttt{Windows~10} vs. \texttt{Windows~11}) are less severe than errors between distant families (e.g., \texttt{Windows} vs. \texttt{Android}).

Alternative hierarchies may also be constructed depending on the operational requirements (see~\Cref{fig:os_tree}), for example, grouping by device type (e.g., \texttt{desktop}, \texttt{server}, or \texttt{mobile}). Note that such groupings extend beyond strict OS fingerprinting and move towards other forms of fingerprinting.

In most works and current deployments, OS fingerprinting is posed as a point prediction task in which a classifier emits a single label (often the maximum a posteriori class) with an optional confidence score. However, such outputs are brittle under class imbalance and distributional shift, and they fail to reflect that error costs depend on the hierarchy (e.g., confusing adjacent \texttt{Windows} versions is typically less severe than confusing \texttt{Windows} with \texttt{Android})~\cite{lastovicka_passive_2023}.

In this context, set-valued prediction via CP~\cite{angelopoulos_conformal_2023} could address these limitations. First, operational decisions benefit from selective prediction: CP exposes an explicit coverage-set-size trade-off, allowing abstention when sets are large (low confidence) and thereby supporting risk-aware policies (e.g., trigger active probing or defer to manual inspection). Second, CP composes seamlessly with any feature pipeline (TCP/IP, DNS, TLS), any model class (trees, neural networks), and with hierarchical decision schemes (as formalised next), serving as a principled uncertainty layer atop existing OS fingerprinting systems. Third, the task is multiclass (hierarchical) with severe class imbalance and non-stationarity: new OS versions appear and prevalence shifts across networks. \Cref{fig:osfing_vs_cp_osfing} illustrates the difference between traditional OS fingerprinting and OS fingerprinting with CP.

CP provides post hoc calibrated, set-valued outputs that remain valid even when the base classifier is misspecified, and it can be periodically recalibrated on fresh traffic from the target environment to track distributional drift. Furthermore, although CP has been applied to various security-related tasks~\cite{yumlembam_comprehensive_2024, papadopoulos_android_2018}, to our knowledge there is no prior literature on its application to either flat (classical) or hierarchical OS fingerprinting.

This provides actionable uncertainty for operations:

\begin{itemize}
	\item in network access control or BYOD settings, a small family-level set may suffice for admission while finer identification proceeds;
	\item in vulnerability management, a focused leaf-level set can prioritise urgent patching, whereas broader sets trigger provisional mitigations; and
	\item in Security Operations Center (SOC) triage, larger sets can raise investigation priority or request additional telemetry, while smaller sets enable automated containment.
\end{itemize}

Additionally, because calibration is lightweight, prediction sets can be periodically recalibrated on fresh traffic from the target environment to track drift without retraining the base models, and selective prediction with risk-coverage controls exposes an explicit accuracy-reliability trade-off for cost-aware decisions in practice~\cite{lastovicka_passive_2023,angelopoulos_conformal_2023}.

Two schemes to equip hierarchical OS fingerprinting with CP, which will be detailed in \Cref{subsec:cp_methods}, are:

\begin{itemize}
	\item \emph{level-wise CP (L-CP)}~\cite{mortier_conformal_2025}, applying CP independently at each hierarchy level and yielding separate prediction sets for each level's labels. This approach is illustrated in~\Cref{fig:l_cp}. While this provides marginal coverage per level, the resulting sets need not be hierarchically consistent (e.g., a child without its parent, or a parent with no admissible children). Moreover, they do not exploit information contained in models trained for finer levels.
	\item \emph{projection-based CP (P-CP)}~\cite{zhang_conformal_2025}, employing CP only at the leaves and then lifting the resulting set to coarser levels by including all ancestors of any retained leaf. This approach is illustrated in~\Cref{fig:p_cp}. It guarantees nested, coherent sets and inherits leaf-level coverage at higher levels. However, it relies exclusively on the leaf predictor, which can inflate coarser-level sets when the leaf model is weak or conservatively calibrated. Furthermore, it ignores strong signals available to dedicated coarse-level models.
\end{itemize}

\section{Materials \& Methods}
\label{sec:materials_methods}

This section details the experimental setup employed to evaluate the proposed conformal prediction schemes. We describe the dataset used for evaluation, the base OS fingerprinting models trained for each hierarchy level, the implementation of both CP methods, and the comprehensive set of evaluation metrics designed to assess validity, efficiency, and structural consistency. Additionally, we provide information on the computational resources utilised throughout this research.

\subsection{Dataset}
\label{subsec:dataset}

\begin{table*}[t]
	\centering
	\caption{Overview of the dataset employed in this study.}
	\vspace{1em}
	\label{tab:dataset_overview}
	\setlength{\tabcolsep}{4pt}
	\renewcommand{\arraystretch}{0.6}
	\begin{tabularx}{\textwidth}
		{>{\centering\arraybackslash}m{0.6cm}
		 >{\centering\arraybackslash}m{3.7cm}
		 >{\centering\arraybackslash}m{1cm}
		 *{6}{>{\centering\arraybackslash}m{0.9cm}}
		 >{\centering\arraybackslash}m{1.2cm}
		 >{\centering\arraybackslash}m{1.7cm}
		 >{\centering\arraybackslash}m{1.5cm}}
		\toprule
		\multirow{2}{*}{\textbf{Year}} &
		\multirow{2}{*}{\textbf{Works}} &
		\multirow{2}{*}{\shortstack{\textbf{Data}\\\textbf{Type}}} &
		\multicolumn{6}{c}{\shortstack{\textbf{Feature Count}}} &
		\multirow{2}{*}{\shortstack{\textbf{Row}\\\textbf{Count}}} &
		\multirow{2}{*}{\shortstack{\textbf{Granularity}}} &
		\multirow{2}{*}{\shortstack{\textbf{Classes}\\\textbf{Count}}}\\
		\cmidrule(lr){4-9}
		& & & \textbf{Total} & \shortstack{\textbf{TCP/IP}} & \textbf{DNS} & \textbf{HTTP} & \textbf{TLS} & \textbf{Other} & & & \\
		\midrule
		2023 &
		\shortstack{\cite{lastovicka_passive_2023}\\\cite{hulak_evaluation_2023}\\\cite{perez-joveApplicationTabularTransformer2025}} &
		IPFIX & 112 & 35 & -- & 7 & 28 & 46 & 109{,}663 &
		\shortstack{\texttt{family}\\\texttt{major}\\\texttt{minor}} &
		\shortstack{12\\50\\88}\\
		\bottomrule
	\end{tabularx}
\end{table*}

For the evaluation of the proposed methods, we utilise the dataset from the study by Lastovicka et al.~\cite{lastovicka_dataset_2023}. This dataset is a recent benchmark for passive OS fingerprinting that captures realistic encrypted web traffic from a university network environment. We selected this specific dataset because, to the best of our knowledge, it is the only one found in the literature that provides the label granularity required for this work, offering ground truth at three hierarchical levels (\texttt{family}, \texttt{major} and \texttt{minor version}). Additionally, it exhibits data characteristics suitable for rigorous experimentation, including a sufficient number of examples per class and a comprehensive set of network traffic features. \Cref{tab:dataset_overview} summarises the main characteristics relevant to our experiments.

The dataset consists of 109,663 flow records collected over an eight-hour period from five servers hosting 475 domains at Masaryk University. The traffic includes connections from a diverse range of devices, including user workstations, mobile phones, and web crawlers. The ground-truth OS labels were derived from HTTP User-Agent strings found in the web server logs, which were then cross-referenced with the corresponding network flows. While this heuristic labelling approach is standard in the field, it is worth noting that it may contain residual noise inherent to User-Agent parsing.

Each record is described by 112 features extracted from multiple protocol layers, including IP and TCP header parameters (35 features), TLS handshake attributes (28 features), and HTTP metadata (7 features), alongside other flow statistics. The dataset supports hierarchical classification with three levels of granularity:

\begin{itemize}
    \item \texttt{family}: 12 broad categories (e.g., \texttt{Windows}, \texttt{Linux}, \texttt{Android}, \texttt{iOS}).
    \item \texttt{major version}: 50 distinct versions (e.g., \texttt{Windows 10}, \texttt{Android 11}).
    \item \texttt{minor version}: 88 fine-grained labels (e.g., \texttt{Windows 10 22H2}).
\end{itemize}

As is typical in real-world network traffic, the class distribution is significantly imbalanced, with popular OSs dominating the sample space. This characteristic makes the dataset particularly suitable for evaluating the robustness of conformal prediction methods, which are designed to provide valid coverage regardless of the underlying data distribution.

\subsection{Base OS Fingerprinting Models}
\label{subsec:base_OS_fing_models}

We trained three independent base models, one for each hierarchical level: OS \texttt{family}, \texttt{major version}, and \texttt{minor version}. For the most granular level, we adapted the classification task to handle OSs that lack a defined \texttt{minor} version in our dataset. Specifically, this model predicts the leaf nodes of the OS tree, where a leaf corresponds either to a concrete \texttt{minor} version or to a generic ``empty'' node representing the major version when no finer distinction is available.

The data preprocessing pipeline handles categorical and numerical features distinctly. Non-byte categorical features are one-hot encoded (with missing values treated as a constant), while byte-based features use ordinal encoding. Numerical features are imputed with the median and normalised using a standard scaler. Feature selection is applied by first removing constant features (variance thresholding) and then selecting the top $k$ features based on Mutual Information, where $k$ is a hyperparameter.

The core predictor employed was a Feed-Forward Neural Network (Multi-Layer Perceptron - MLP). This architecture was selected because it is a standard and effective choice for tabular data classification tasks and, crucially, it naturally outputs the class logits required to compute the non-conformity scores for conformal prediction.

The MLP architecture consists of a sequence of linear layers, each followed by Batch Normalization and a Rectified Linear Unit (ReLU) activation function. Dropout is applied after activations to prevent overfitting. The network depth and width are determined by hyperparameters, typically ranging from smaller networks (e.g., [64, 32]) for the family level to deeper, wider networks (e.g., [1024, 512]) for the leaf level.

Training is performed using the AdamW optimizer and a weighted Cross Entropy loss function, which penalises errors on minority classes more heavily (weights are inversely proportional to class frequency). To further address class imbalance, the pipeline employs random undersampling of majority classes and optionally applies Synthetic Minority Over-sampling Technique (SMOTE) to the training folds. The learning rate is managed by a OneCycleLR scheduler, and training includes early stopping based on validation F1-score improvement.

Hyperparameters were selected to maximise the validation Macro F1-Score through a random grid search validated via stratified \(K\)-fold cross-validation with \(K = 5\). For every configuration in the search grid, we perform 5-fold stratified cross-validation, select the configuration with the highest mean Macro F1-Score, and then retrain the final model from scratch on the full training dataset.

It is important to remark that the specific architecture or performance of the base classifiers is not the primary focus of this work, but rather the CP schemes applied to them. These networks serve merely as representative base models: both CP schemes proposed in this work can, in principle, be applied to any OS fingerprinting classifier that provides logits (or scores) that can be interpreted as a predictive probability distribution over classes.

\subsection{Conformal Prediction Methods}
\label{subsec:cp_methods}

This subsection formalises the two schemes utilised in this work for applying CP to OS fingerprinting over a hierarchical label space. We first introduce notation and the learning/calibration setup, and then detail: (i) \emph{level-wise CP} (L-CP), applying CP separately at each level, and (ii) \emph{projection-based CP} (P-CP), applying CP at the leaves only and then projecting sets to upper levels.

\subsubsection{Notation and Setup}
\label{subsubsec:baseline_notation}

Let $\mathcal{X}$ denote the instance space (e.g., flow-/packet-derived features). The label space is organised as a rooted tree with $K$ levels, from coarsest ($k=1$) to finest ($k=K$). Our mathematical formulation explicitly allows any $K$ by modelling parent–child maps on a rooted tree; it remains applicable irrespective of how the OS labels are constructed or collapsed in a given study or dataset.

The hierarchy is generally incomplete: not all nodes at intermediate levels have children. Let $\mathcal{Y}_k=\{1,\dots,m_k\}$ be the (possibly empty) set of nodes at level $k$. For $k<K$, let $\mathrm{Ch}(y_k)\subseteq \mathcal{Y}_{k+1}$ denote the set of children of $y_k$, and set $\mathrm{Ch}(y_K)=\varnothing$ by convention.

\begin{equation}
	\mathrm{parent}:\bigcup_{k=1}^{K-1}\mathcal{Y}_{k+1}\longrightarrow \bigcup_{k=1}^{K-1}\mathcal{Y}_k
	\label{eq:baseline-notation-parent}
\end{equation}

\noindent where for $y\in\mathcal{Y}_{k+1}$ we have $\mathrm{parent}(y)\in\mathcal{Y}_k$, and $\mathrm{parent}^{-1}(y_k)=\mathrm{Ch}(y_k)$.

Define the set of terminal leaves

\begin{equation}
	\mathcal{L}\triangleq \bigl\{\, y \in \bigcup_{k=1}^{K}\mathcal{Y}_k \;:\; \mathrm{Ch}(y)=\varnothing \,\bigr\}.
	\label{eq:baseline-notation-termleaves}
\end{equation}

Write $y'\succeq y$ if $y'$ is a (transitive) descendant of $y$, and define the terminal–descendant operator

\begin{equation}
	\mathrm{LeafDesc}(y_k) \triangleq \bigl\{\, y \in \mathcal{L} \;:\; y \succeq y_k \,\bigr\}.
	\label{eq:baseline-notation-leafdesc}
\end{equation}

For convenience, index the leaf layer abstractly as $\mathcal{Y}_{\mathrm{leaf}}\equiv \mathcal{L}$ (even if branches have different depths) and let $\pi_{k\leftarrow \mathrm{leaf}}:\mathcal{Y}_{\mathrm{leaf}}\to \mathcal{Y}_k$ denote the ancestor map whenever the level-$k$ ancestor exists.

We assume exchangeable data $\{(X_i,Y_{\mathrm{leaf},i})\}_{i=1}^n$ with $Y_{\mathrm{leaf},i}\in\mathcal{Y}_{\mathrm{leaf}}$, and define induced ancestors $Y_{k,i}=\pi_{k\leftarrow \mathrm{leaf}}(Y_{\mathrm{leaf},i})$ whenever the level-$k$ ancestor exists.

For each level $k$ with $m_k>0$, suppose we have trained a probabilistic classifier $f_k:\mathcal{X}\to\Delta^{m_k-1}$ (e.g., softmax), yielding scores $p_k(y\mid x)$ for $y\in\mathcal{Y}_k$. Define the nonconformity score (smaller is “more conforming”)

\begin{equation}
	s_k(x,y) \triangleq 1- p_k(y\mid x),
	\label{eq:baseline-notation-nonconformity-score}
\end{equation}

\noindent though any monotone surrogate (e.g., $-\log p_k(y\mid x)$) may be used. For split CP, partition the data into a proper training set (to fit $f_k$) and, for each level $k$, a calibration set $\mathcal{C}_k=\{(X_i,Y_{k,i})\}_{i=1}^{n_k}$ containing only samples for which the level-$k$ ancestor $Y_{k,i}$ is defined. With calibration scores $A^k_i \triangleq s_k(X_i,Y_{k,i})$, define the conservative empirical quantile

\begin{equation}
	\widehat{q}_k(1-\alpha_k) \triangleq \mathrm{Quantile}^{\mathrm{higher}}_{\frac{\lceil (n_k+1)(1-\alpha_k)\rceil}{n_k+1}}\big(\{A^k_i\}_{i=1}^{n_k}\big).
	\label{eq:baseline-notation-q}
\end{equation}

for a target miscoverage $\alpha_k\in(0,1)$ at level $k$.

\subsubsection{Level-wise CP (L-CP)}
\label{subsubsec:lw_cp}

\begin{figure*}[t]
	\centering
	\includegraphics[width=\linewidth]{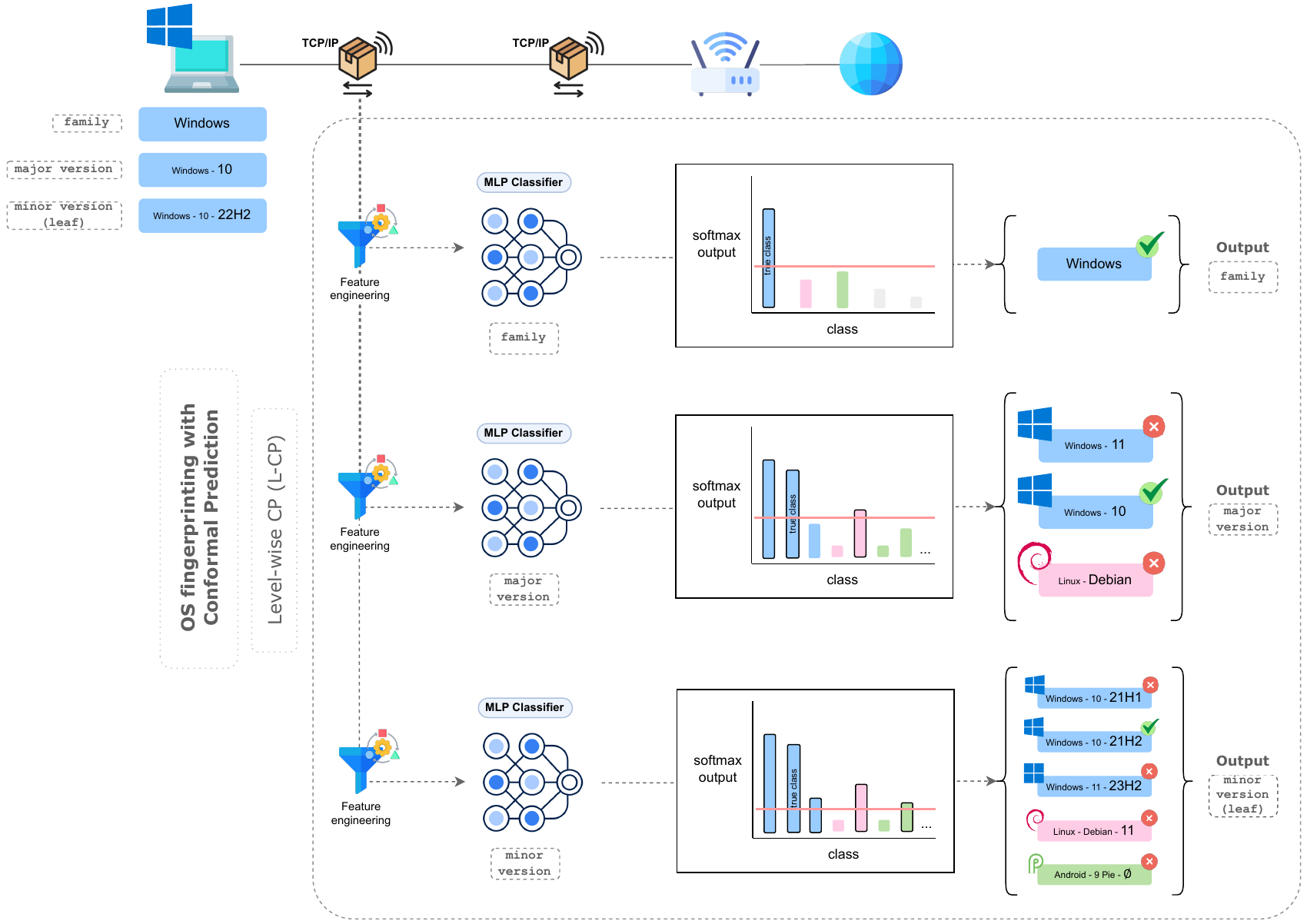}
	\caption{Diagram illustrating the \emph{level-wise CP (L-CP)} method for hierarchical OS fingerprinting. The system processes network traffic through feature engineering and applies separate MLP classifiers at each hierarchy level (\texttt{family}, \texttt{major version}, and \texttt{minor version}). Each classifier independently generates softmax outputs, which are then processed using Conformal Prediction to produce level-specific prediction sets. The figure shows how prediction sets are constructed independently at each level, with separate coverage guarantees, but without enforcing hierarchical consistency across levels.}
	\label{fig:l_cp}
\end{figure*}

For a test instance $x$, define the level-$k$ prediction set by applying CP independently at each level (see \Cref{fig:l_cp}):

\begin{equation}
	\Gamma^{(A)}_{k,\alpha_k}(x) \triangleq \bigl\{\, y\in\mathcal{Y}_k \;:\; s_k(x,y) \le \widehat{q}_k(1-\alpha_k) \,\bigr\}.
	\label{eq:l-CP-levelwise-set}
\end{equation}

Under exchangeability, \eqref{eq:l-CP-levelwise-set} guarantees marginal coverage at each level,

\begin{equation}
	\mathbb{P}\!\left\{\, Y_k \in \Gamma^{(A)}_{k,\alpha_k}(X) \,\middle|\, Y_k \right\} \ge 1-\alpha_k,
	\qquad \text{for all levels } k.
	\label{eq:l-CP-marginal-coverage}
\end{equation}

\paragraph{Limitations} Because sets are constructed independently across levels, they need not be hierarchically consistent. Two failure modes are typical:

\begin{itemize}
	\item (child-without-parent):

	$\exists\, y_{k+1}\in \Gamma^{(A)}_{k+1,\alpha_{k+1}}(x)\ \text{s.t.}\ \mathrm{parent}(y_{k+1}) \notin \Gamma^{(A)}_{k,\alpha_k}(x)$,

	\item (parent-without-child):

	$\exists\, y_k\in \Gamma^{(A)}_{k,\alpha_k}(x)\ \text{s.t.}\ \Gamma^{(A)}_{k+1,\alpha_{k+1}}(x)\cap \mathrm{Ch}(y_k)=\varnothing$.
\end{itemize}

Hence, a decision procedure that requires cross-level coherence (e.g., top-down selection) may face ambiguous or incompatible sets. Moreover, informational inefficiency can arise: $\Gamma^{(A)}_{k,\alpha_k}$ does not leverage scores from finer levels, even when $f_{k+1}$ contains discriminative signal.

\subsubsection{Projection-based CP (P-CP)}
\label{subsubsec:p_cp}

\begin{figure*}[t]
	\centering
	\includegraphics[width=\linewidth]{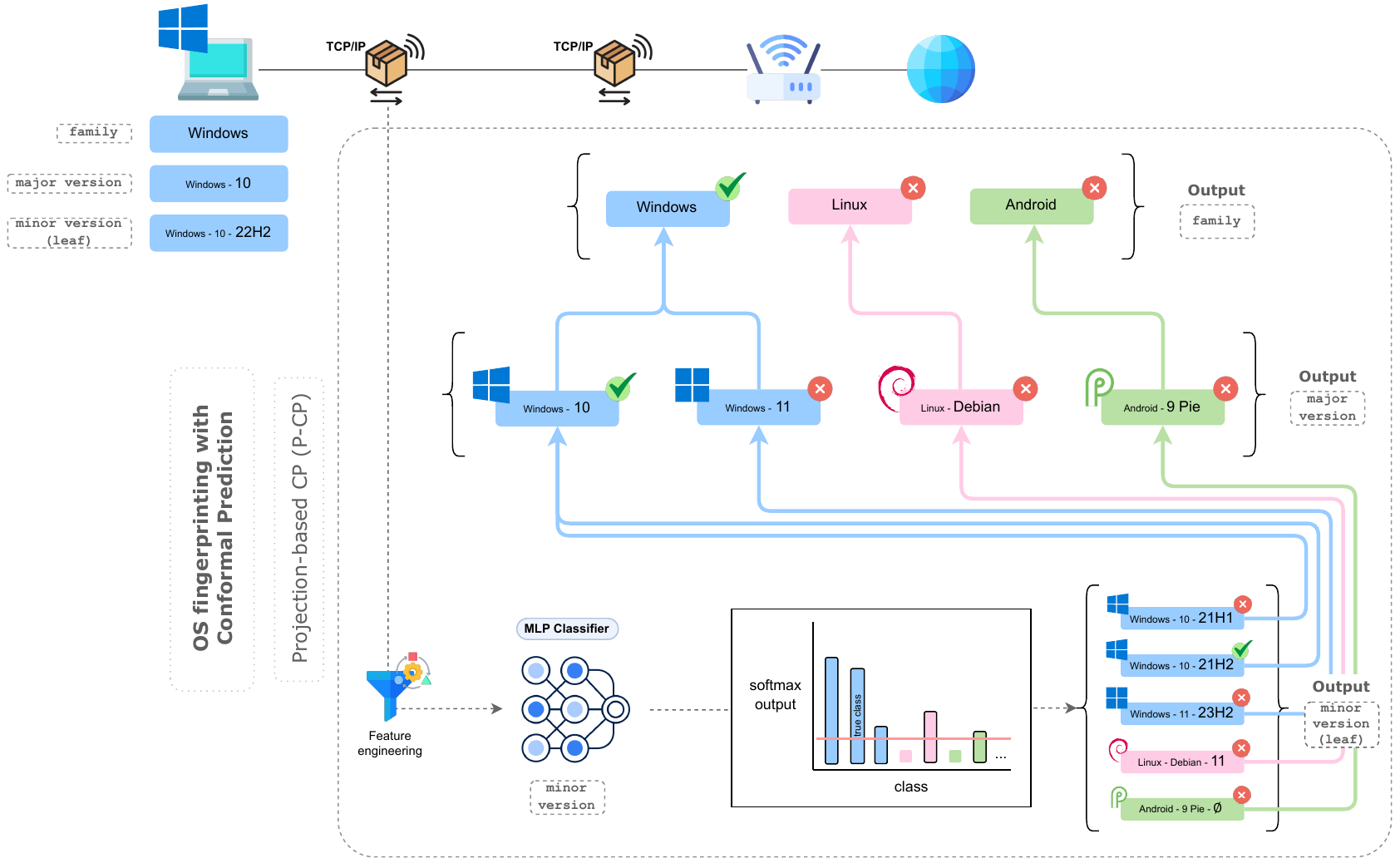}
	\caption{Diagram illustrating the \emph{projection-based CP (P-CP)} method for hierarchical OS fingerprinting. The system applies CP only at the leaf (\texttt{minor version}) level, generating a prediction set based on the MLP classifier's softmax outputs. The leaf-level prediction set is then projected upwards to coarser levels (\texttt{major version} and \texttt{family}) by including all ancestors of any retained leaf. This upward projection ensures hierarchical consistency, as shown by the nested structure where predictions at coarser levels are derived from the leaf-level set, guaranteeing that every predicted class corresponds to a valid path in the OS taxonomy.}
	\label{fig:p_cp}
\end{figure*}

Here CP is applied only at the leaf level ($k=K$), as illustrated in \Cref{fig:p_cp}. Construct the leaf set

\begin{equation}
	\Gamma^{(B)}_{\mathrm{leaf},\alpha}(x) \triangleq
	\bigl\{\, y\in\mathcal{Y}_{\mathrm{leaf}} \;:\; s_{\mathrm{leaf}}(x,y) \le \widehat{q}_{\mathrm{leaf}}(1-\alpha) \,\bigr\}.
	\label{eq:p-CP-leaf-set}
\end{equation}

\noindent with marginal coverage $\mathbb{P}\{Y_{\mathrm{leaf}}\in \Gamma^{(B)}_{\mathrm{leaf},\alpha}(X)\}\ge 1-\alpha$. Then project upwards to coarser levels by including exactly those ancestors that have at least one descendant in the leaf set:

\begin{equation}
	\begin{aligned}
		\Gamma^{(B)}_{k,\alpha}(x) \triangleq\;
		&\bigl\{\, y\in\mathcal{Y}_k \;:\; \mathrm{LeafDesc}(y)\cap
		\Gamma^{(B)}_{\mathrm{leaf},\alpha}(x)\neq \varnothing \,\bigr\},\\
		&\qquad k=1,\dots,K-1.
	\end{aligned}
	\label{eq:p-CP-upward-projection}
\end{equation}

By construction, the sets are hierarchically consistent and nested:

\begin{equation}
	\begin{gathered}
		y_{k+1}\in \Gamma^{(B)}_{k+1,\alpha}(x)\;\Rightarrow\;
		\mathrm{parent}(y_{k+1})\in \Gamma^{(B)}_{k,\alpha}(x),\\
		\text{and}\ \Gamma^{(B)}_{1,\alpha}(x)\supseteq \cdots \supseteq \Gamma^{(B)}_{K-1,\alpha}(x)\supseteq
		\Gamma^{(B)}_{\mathrm{leaf},\alpha}(x).
	\end{gathered}
	\label{eq:p-CP-hierarchical-consistency}
\end{equation}

Moreover, coverage at coarser levels is inherited from leaf-level coverage. Indeed,

\begin{equation}
	\{Y_{\mathrm{leaf}} \in \Gamma^{(B)}_{\mathrm{leaf},\alpha}(X)\}\ \Rightarrow\ \{\pi_{k\leftarrow \mathrm{leaf}}(Y_{\mathrm{leaf}})\in \Gamma^{(B)}_{k,\alpha}(X)\}.
	\label{eq:p-CP-coverage}
\end{equation}

\noindent so that

\begin{equation}
	\mathbb{P}\{Y_k\in \Gamma^{(B)}_{k,\alpha}(X)\}\ge 1-\alpha,\qquad \text{for all levels } k \text{ with } Y_k \text{ defined.}
	\label{eq:p-CP-inheritance}
\end{equation}

\noindent hence $\mathbb{P}\{Y_k\in \Gamma^{(B)}_{k,\alpha}(X)\}\ge \mathbb{P}\{Y_{\mathrm{leaf}}\in \Gamma^{(B)}_{\mathrm{leaf},\alpha}(X)\}\ge 1-\alpha$ for all $k<K$.

\paragraph{Limitations} P-CP uses only the leaf classifier $f_K$. If $f_K$ is misspecified or weak for certain leaves, the induced sets at upper levels can be unnecessarily large; conversely, strong signals available to $f_k$ for small $k$ (e.g., family-level cues) are not utilised to tighten $\Gamma^{(B)}_{k,\alpha}$. In addition, because upward projection is a set-theoretic expansion, any calibration conservativeness at the leaves propagates and may be amplified at higher levels (larger set sizes).

\subsubsection{Experimental Implementation Details}
\label{subsubsec:experimental_implementation}

The proposed CP schemes are implemented within a comprehensive framework designed to ensure reproducibility and strict statistical validity.

\paragraph{Data Preparation}
The partition into training (70\%), calibration (15\%), and test (15\%) sets is performed using a stratified split based on \texttt{leaf} classes. This strategy ensures that even rare OS versions are adequately represented in all subsets. Classes with extremely low prevalence (fewer than 2 samples) are handled via random splitting to prevent evaluation artifacts.

\paragraph{Prediction Logic}
For L-CP, the system computes non-conformity scores independently for each level using the corresponding base model's softmax outputs. In the case of P-CP, the implementation strictly enforces consistency by deriving the upper-level sets solely from the valid leaf prediction set. Specifically, the \texttt{major} set is generated as the unique set of parents of all retained leaves, and the \texttt{family} set as the unique set of grandparents, ensuring $\mathrm{HIR}=0$ by construction.

\paragraph{Evaluation Framework}
To account for the variability inherent in random splitting, the evaluation reports aggregated metrics over 50 independent Monte Carlo iterations. The framework systematically sweeps significance levels $\alpha \in [0.0, 0.5]$ to characterise the full coverage-efficiency trade-off curve.

\subsection{Evaluation Metrics}
\label{subsec:eval_metrics}

To assess the validity and efficiency of all the CP approaches across the different hierarchy levels (\texttt{family}, \texttt{major version}, and \texttt{leaf/minor version}), we employ the following metrics evaluated on the test set of size $N_{\mathrm{test}}$.

\paragraph{Coverage}
The primary validity metric measuring the proportion of test samples for which the true label $y_i$ is contained within the prediction set $\Gamma(x_i)$. For a user-specified miscoverage rate $\alpha$, a calibrated conformal predictor should ideally satisfy:

\begin{equation}
	\text{Coverage} = \frac{1}{N_{\mathrm{test}}} \sum_{i=1}^{N_{\mathrm{test}}} \mathbb{I}(y_i \in \Gamma(x_i)) \ge 1 - \alpha
\end{equation}

\paragraph{Set Size}
This metric quantifies the informational efficiency (or granularity) of the predictor. It is defined as the average cardinality of the prediction sets. Smaller sets are preferred, provided that the target coverage is maintained.

\begin{equation}
	\text{Mean Set Size} = \frac{1}{N_{\mathrm{test}}} \sum_{i=1}^{N_{\mathrm{test}}} |\Gamma(x_i)|
\end{equation}

\begin{figure*}[t]
	\centering
	\includegraphics[width=0.8\linewidth]{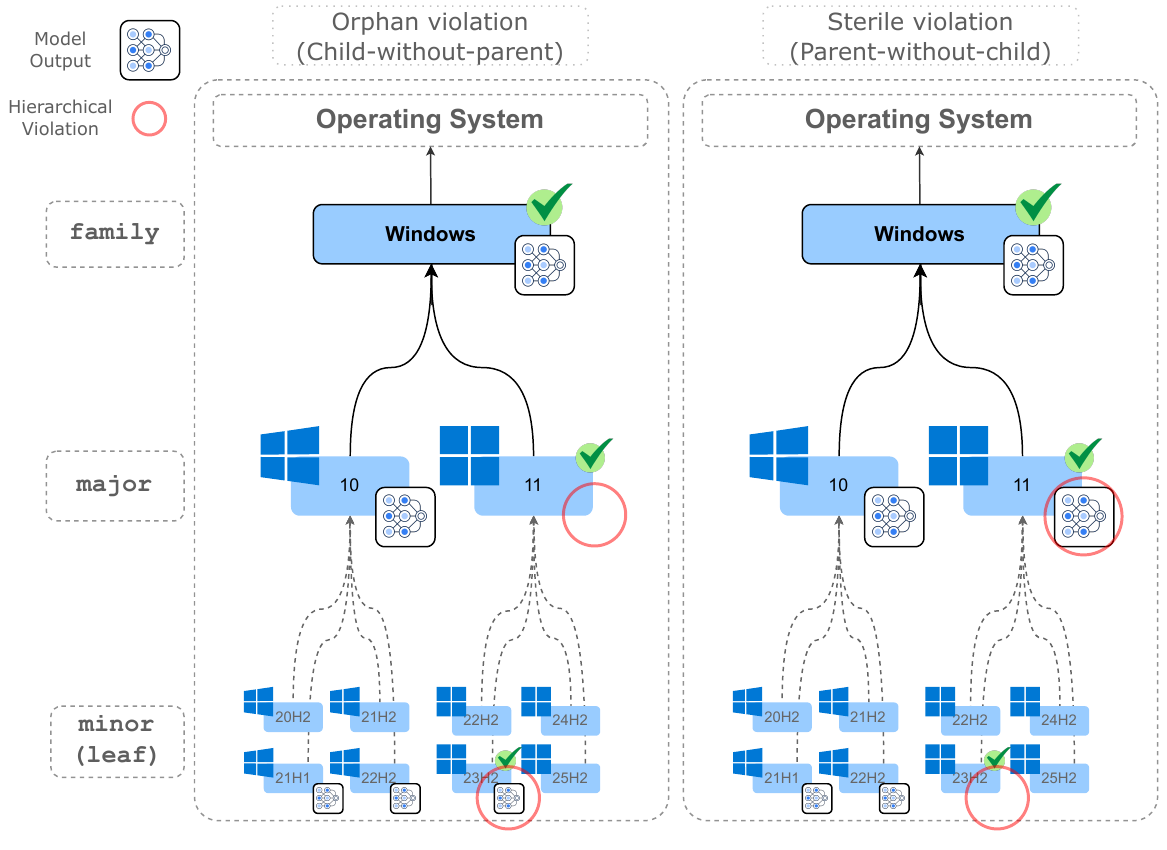}
	\caption{Visual illustration of the two types of hierarchical violations that contribute to the Hierarchical Inconsistency Rate (HIR). The left diagram demonstrates an \emph{orphan violation}, where a child node (\texttt{Windows 11 - 23H2}) is predicted without its parent (\texttt{Windows 11}) being in the prediction set. The right diagram shows a \emph{sterile violation}, where a parent node (\texttt{Windows 11}) is predicted but none of its children are included in the prediction set. These violations indicate structural inconsistencies in hierarchical predictions that can undermine the reliability of automated decision-making systems.}
	\label{fig:hir_violations}
\end{figure*}

\paragraph{Hierarchical Inconsistency Rate (HIR)}
Beyond standard CP metrics, it is crucial to evaluate whether the predicted sets at different granularity levels form a coherent path in the OS taxonomy. We define two types of structural inconsistencies between a parent level $k$ and a child level $k+1$ (see \Cref{fig:hir_violations}):

\begin{enumerate}
	\item \emph{Orphan violation (Child-without-parent):} A specific OS version is predicted at level $k+1$, but its corresponding parent family or version is not present in the prediction set at level $k$.
	\item \emph{Sterile violation (Parent-without-child):} A node is predicted at level $k$ (and is not a taxonomic leaf), but none of its valid children appear in the prediction set at level $k+1$.
\end{enumerate}

These two types of violations are visually illustrated in \Cref{fig:hir_violations} using the Windows OS hierarchy as an example.

Formally, for a test input $x$, we define the inconsistency indicator $V(x)$ as:

\begin{equation}
	V(x) = \mathbb{I}\left( \bigcup_{k=1}^{K-1} \left( \mathcal{E}^{\mathrm{orphan}}_{k}(x) \cup \mathcal{E}^{\mathrm{sterile}}_{k}(x) \right) \neq \varnothing \right)
\end{equation}

where the sets of violating edges are defined as:

\begin{gather}
	\begin{aligned}
		\mathcal{E}^{\mathrm{orphan}}_{k}(x)
		&= \left\{ y \in \Gamma_{k+1}(x) : \mathrm{parent}(y) \notin \Gamma_{k}(x) \right\},
	\end{aligned}\\
	\begin{aligned}
		\mathcal{E}^{\mathrm{sterile}}_{k}(x)
		&= \left\{ y \in \Gamma_{k}(x) : \mathrm{Ch}(y) \neq \varnothing \land
		\Gamma_{k+1}(x) \cap \mathrm{Ch}(y) = \varnothing \right\}.
	\end{aligned}
\end{gather}

The HIR is finally calculated as the proportion of test samples containing at least one violation across any level of the hierarchy:

\begin{equation}
	\mathrm{HIR} = \frac{1}{N_{\mathrm{test}}} \sum_{i=1}^{N_{\mathrm{test}}} V(x_i).
\end{equation}

A value of $0$ indicates perfect structural coherence, while higher values indicate that the operator receives contradictory information from the conformal predictor.

\paragraph{Empty Set Rate}
The proportion of test samples for which the model returns an empty prediction set ($\Gamma(x_i) = \varnothing$). In the standard inductive conformal prediction setting, empty sets generally indicate that the test sample is significantly different from the calibration distribution (out-of-distribution) or that the model is highly uncertain.

\begin{equation}
	\text{Empty Rate} = \frac{1}{N_{\mathrm{test}}} \sum_{i=1}^{N_{\mathrm{test}}} \mathbb{I}(|\Gamma(x_i)| = 0)
\end{equation}

\paragraph{Singleton Rate}
The proportion of test samples for which the prediction set contains exactly one label. This metric serves as a proxy for ``usability''; a high singleton rate implies that the conformal predictor often yields precise, unambiguous predictions similar to a standard point classifier, but with statistical validity.
\begin{equation}
	\text{Singleton Rate} = \frac{1}{N_{\mathrm{test}}} \sum_{i=1}^{N_{\mathrm{test}}} \mathbb{I}(|\Gamma(x_i)| = 1)
\end{equation}

\subsection{Computational Resources}
\label{subsec:computational_resources}

Model development and training primarily leveraged the FinisTerrae III supercomputer~\cite{cesga_finisterrae}. This Bull ATOS bullx cluster (4.36 PetaFLOPS peak) comprises 714 Intel Xeon Ice Lake 8352Y processors, 157 GPUs (141 Nvidia A100 and 16 T4), 126 TB of memory, 359 TB of NVMe storage, and Infiniband HDR 100 networking, enabling accelerated parallel experimentation. Resources were allocated dynamically based on task requirements.

To ensure reproducibility, experiments were validated on a standard HP EliteDesk 800 G6 workstation (Windows 11 Pro, Intel 2.90 GHz CPU, 64 GB RAM). This confirms the pipeline runs effectively on commercial hardware. Notably, the proposed CP schemes add negligible computational overhead during calibration and testing.

\section{Results \& Discussion}
\label{sec:results_discussion}

This section presents a comprehensive evaluation of the two CP schemes across multiple significance levels, examining their validity, efficiency, and structural consistency. We analyse the empirical results to understand the trade-offs between level-wise efficiency and structural consistency, and discuss the operational implications for different deployment scenarios in network security applications.

\subsection{Results}
\label{subsec:results}

\begin{figure*}[!ht]
	\centering
	\includegraphics[width=0.68\linewidth]{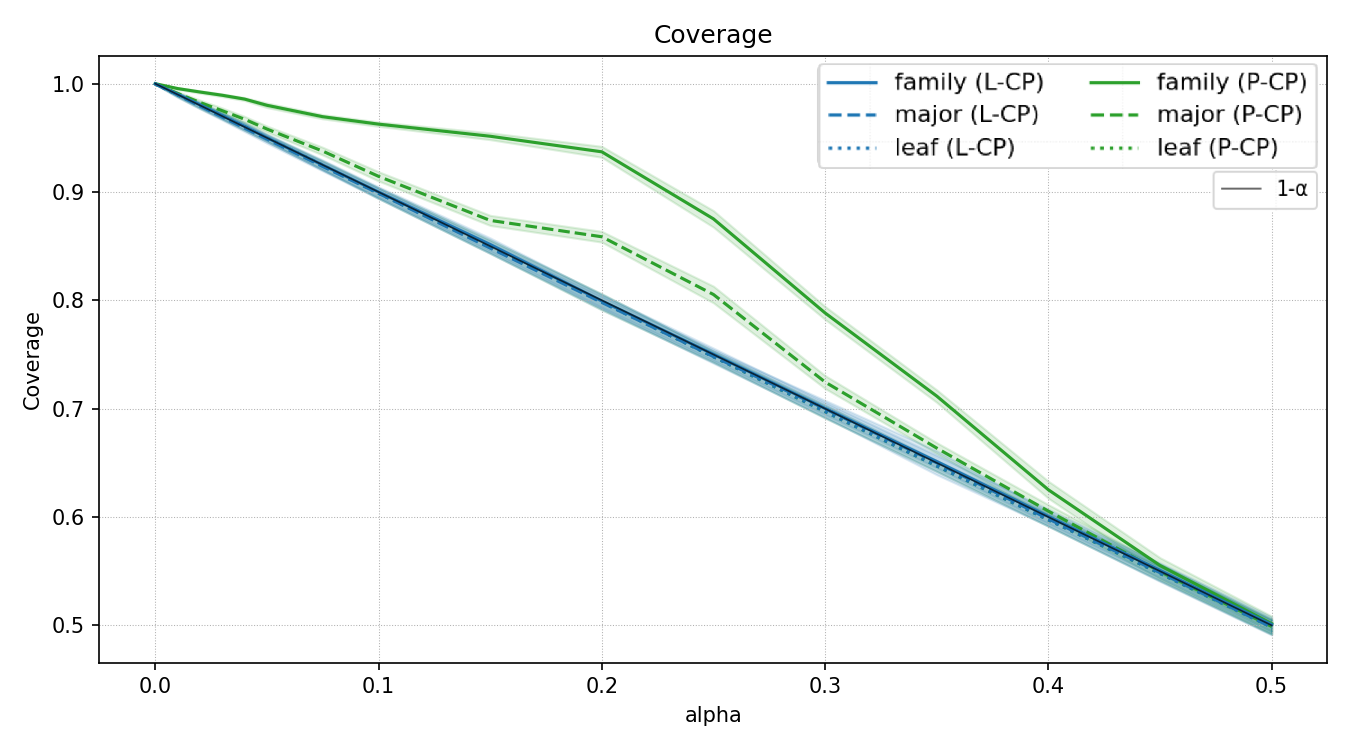}
\caption{Empirical marginal coverage across hierarchy levels. The plot compares the coverage of L-CP and P-CP against the nominal target $1-\alpha$. }
	\label{fig:coverage}
\end{figure*}

\begin{figure*}[!ht]
	\centering
	\includegraphics[width=0.68\linewidth]{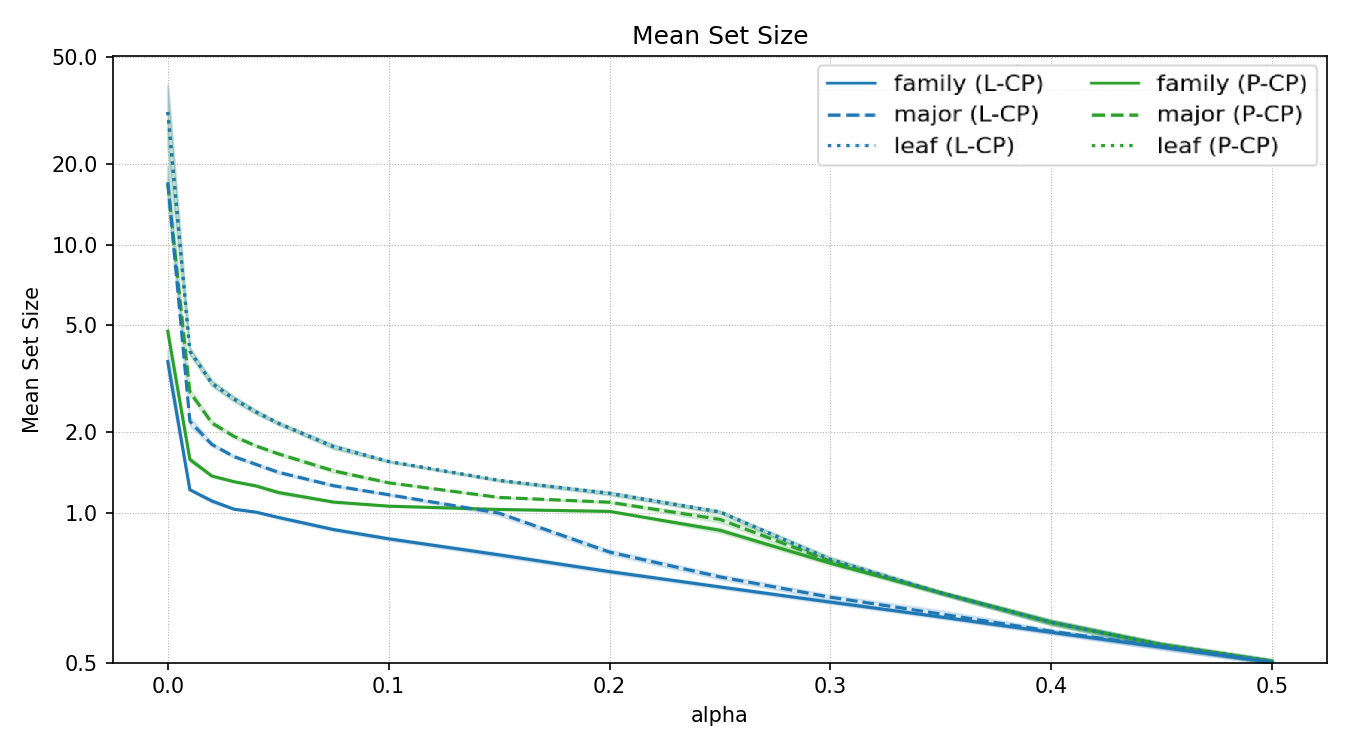}
	\caption{Prediction efficiency measured by mean set size. The figure shows the average size of prediction sets for different $\alpha$ values.}
	\label{fig:set_size}
\end{figure*}

\begin{figure*}[!ht]
	\centering
	\includegraphics[width=0.68\linewidth]{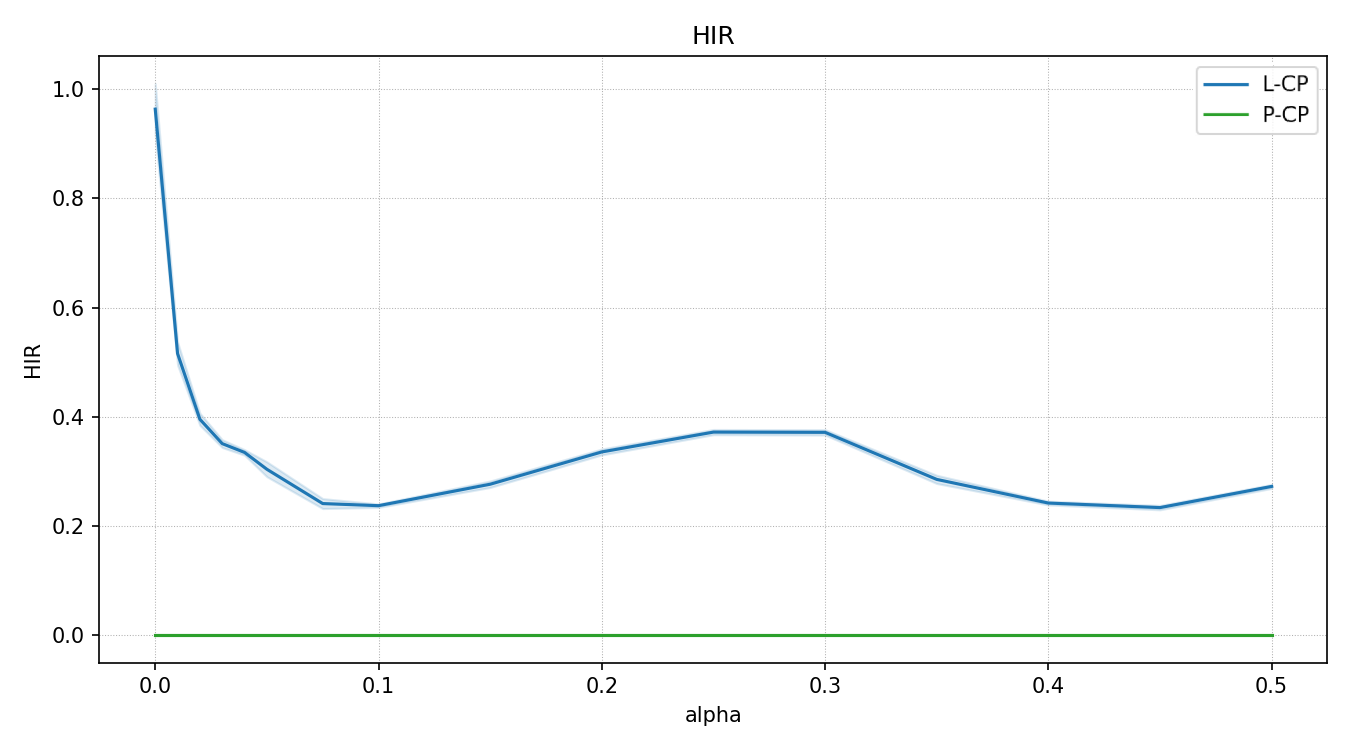}
	\caption{HIR comparison. The metric quantifies the frequency of taxonomic violations (e.g., orphan or sterile sets).}
	\label{fig:hir}
\end{figure*}

\begin{figure*}[!ht]
	\centering
	\includegraphics[width=0.68\linewidth]{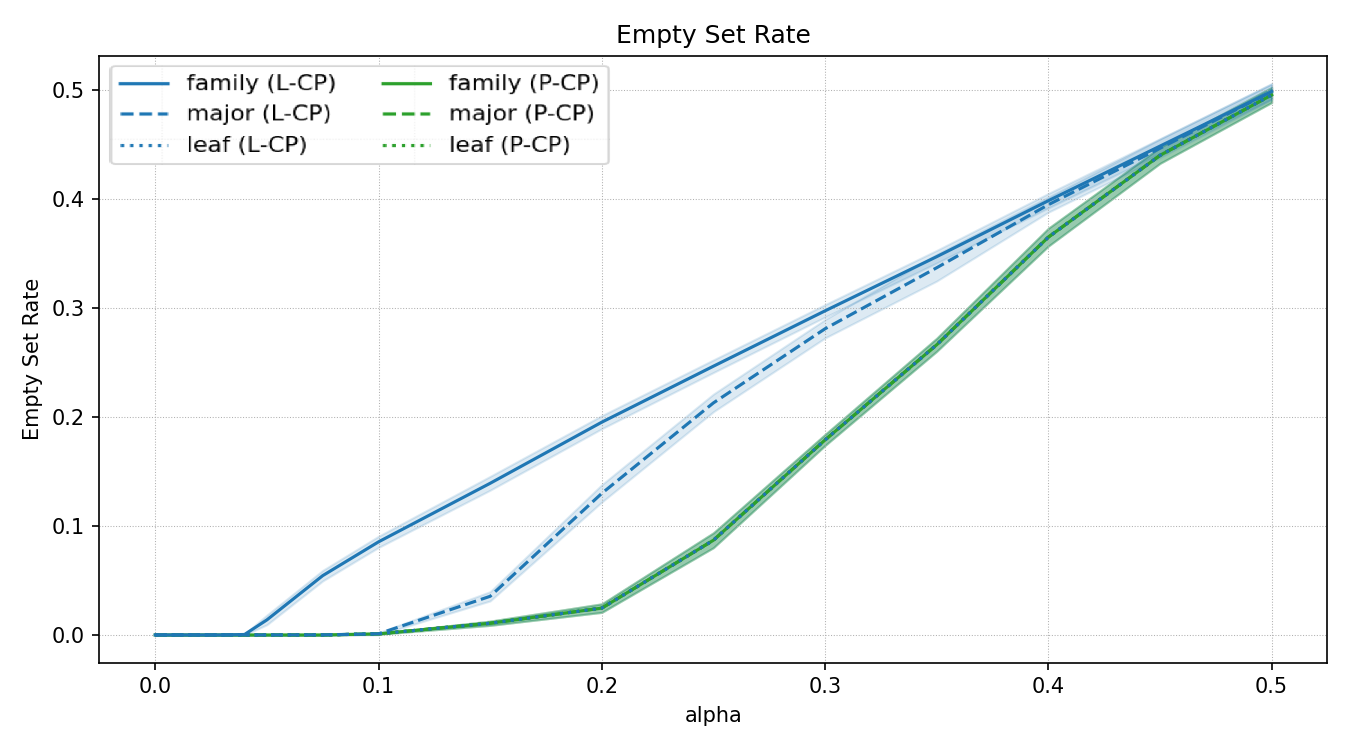}
	\caption{Frequency of empty prediction sets. The plot shows the rate at which methods abstain from prediction (return $\varnothing$) as $\alpha$ increases.}
	\label{fig:empty_rate}
\end{figure*}

\begin{figure*}[!ht]
	\centering
	\includegraphics[width=0.68\linewidth]{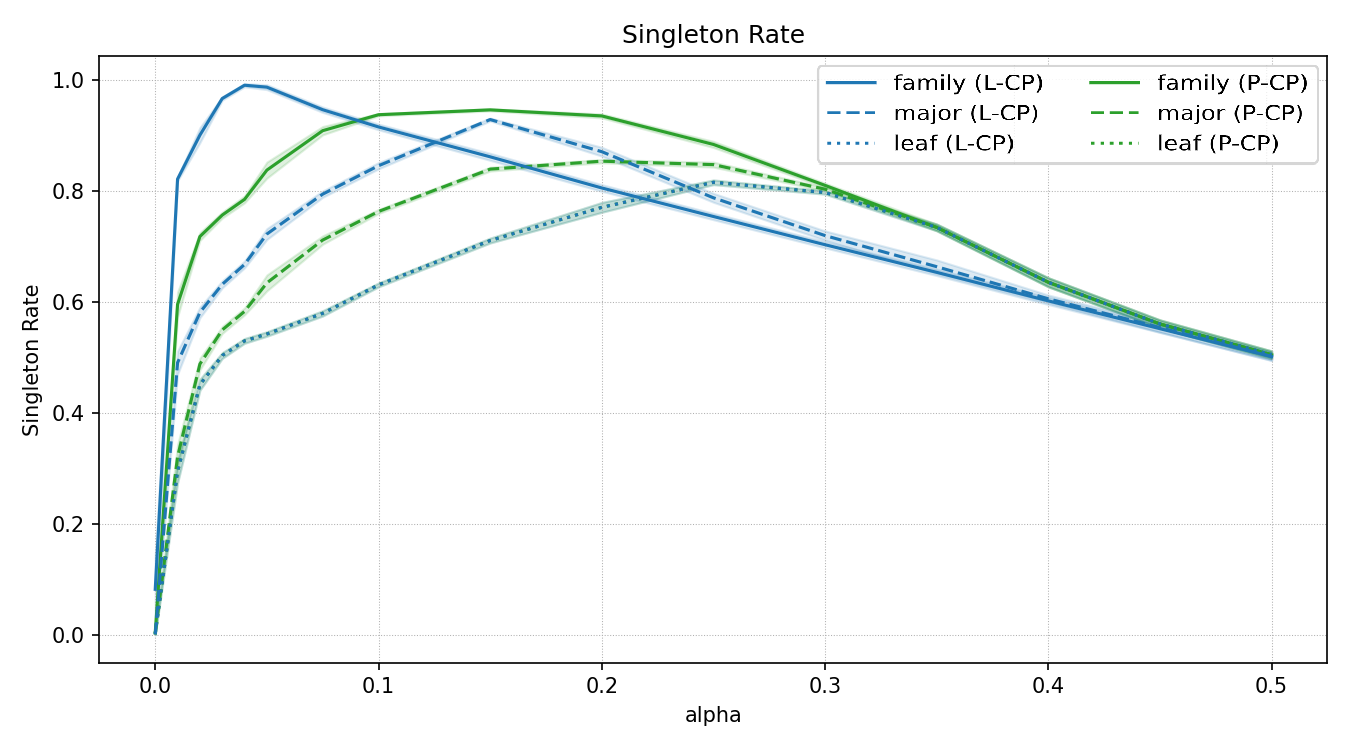}
	\caption{Singleton rate (usability) comparison. This metric represents the proportion of samples where the prediction set contains exactly one label.}
	\label{fig:singleton_rate}
\end{figure*}

We evaluate the two proposed CP schemes, \emph{level-wise CP (L-CP)} and \emph{projection-based CP (P-CP)}, across a range of significance levels $\alpha \in [0, 0.5]$. The results, aggregated over 50 independent runs, are summarised in \Cref{tab:cp_results} and visualised in \Cref{fig:coverage,fig:set_size,fig:hir,fig:empty_rate,fig:singleton_rate}. To facilitate reproducibility, all the code to reproduce the experiments is available under the GNU GPL v3 License in the following public GitHub repository: \RepoLink.

\clearpage

\clearpage

\begin{table*}[!ht]
	\centering
\caption{Quantitative comparison of L-CP and P-CP approaches for significance levels $\alpha = 0.02$, $0.05$, $0.10$, and $0.20$. Results are reported as $\mathrm{mean} \pm \mathrm{std}$ over 50 independent runs.}
	\label{tab:cp_results}
	\vspace{1em}
		\begin{tabular}{ccccccc}
			\toprule
			& & \multicolumn{2}{c}{\textbf{Family Level}} & \multicolumn{2}{c}{\textbf{Leaf Level}} & \textbf{Structural} \\
			\cmidrule(lr){3-4} \cmidrule(lr){5-6} \cmidrule(lr){7-7}
			Target $\alpha$ & Method & Coverage $\uparrow$ & Mean Set Size $\downarrow$ & Coverage $\uparrow$ & Mean Set Size $\downarrow$ & HIR $\downarrow$ \\
			\midrule

			\multirow{2}{*}{0.02}
			& L-CP & $0.980 \pm 0.002$ & $\mathbf{1.111 \pm 0.015}$ & $0.980 \pm 0.003$ & $3.046 \pm 0.071$ & $0.395 \pm 0.011$ \\
			& P-CP & $\mathbf{0.992 \pm 0.001}$ & $1.375 \pm 0.014$ & $0.980 \pm 0.003$ & $3.046 \pm 0.071$ & $\mathbf{0.000 \pm 0.000}$ \\
			\midrule

			\multirow{2}{*}{0.05}
			& L-CP & $0.950 \pm 0.003$ & $\mathbf{0.986 \pm 0.004}$ & $0.950 \pm 0.003$ & $2.163 \pm 0.026$ & $0.304 \pm 0.014$ \\
			& P-CP & $\mathbf{0.980 \pm 0.002}$ & $1.195 \pm 0.016$ & $0.950 \pm 0.003$ & $2.163 \pm 0.026$ & $\mathbf{0.000 \pm 0.000}$ \\
			\midrule

			\multirow{2}{*}{0.10}
			& L-CP & $0.900 \pm 0.005$ & $\mathbf{0.915 \pm 0.005}$ & $0.899 \pm 0.005$ & $1.559 \pm 0.012$ & $0.237 \pm 0.003$ \\
			& P-CP & $\mathbf{0.963 \pm 0.002}$ & $1.063 \pm 0.002$ & $0.899 \pm 0.005$ & $1.559 \pm 0.012$ & $\mathbf{0.000 \pm 0.000}$ \\
			\midrule

			\multirow{2}{*}{0.20}
			& L-CP & $0.799 \pm 0.006$ & $\mathbf{0.805 \pm 0.006}$ & $0.799 \pm 0.007$ & $1.186 \pm 0.017$ & $0.336 \pm 0.005$ \\
			& P-CP & $\mathbf{0.937 \pm 0.005}$ & $1.016 \pm 0.005$ & $0.799 \pm 0.007$ & $1.186 \pm 0.017$ & $\mathbf{0.000 \pm 0.000}$ \\

			\bottomrule
		\end{tabular}
\end{table*}

\clearpage

\paragraph{Validity (Coverage)}
Both methods satisfy the marginal coverage guarantee. L-CP achieves coverage rates very close to the nominal target $1-\alpha$ at all three hierarchy levels (family, major, leaf), as shown in \Cref{fig:coverage}. This confirms that the level-wise calibration is effective locally (e.g., at $\alpha=0.10$, L-CP family coverage is $90.0\%$). In contrast, P-CP, which calibrates only at the leaves and projects sets upwards, exhibits systematic over-coverage at coarser levels. For example, at $\alpha=0.10$ (target 90\%), P-CP achieves a family-level coverage of $96.3\%$. This conservativeness arises because the family-level set is constructed as the union of the ancestors of all valid leaves; if any leaf descendant is plausible, the family must be included, often retaining families that a dedicated family-level classifier might have confidently rejected (e.g., retaining the whole \texttt{Windows} family because a single noisy \texttt{Windows 7} leaf prediction remains valid).

\paragraph{Efficiency (Set Size)}
The trade-off for P-CP's structural consistency is reduced efficiency at coarser levels. \Cref{fig:set_size} illustrates that L-CP generally produces smaller prediction sets at the family level (e.g., mean size $\approx 0.92$ at $\alpha=0.10$) compared to P-CP ($\approx 1.06$). L-CP's mean set size drops below 1.0 at moderate $\alpha$ due to the increasing frequency of empty sets (see \Cref{fig:empty_rate}), a behaviour permitted by the conformal framework to signal out-of-distribution samples or high uncertainty. P-CP produces fewer empty sets at the family level because a set is empty only if \emph{all} leaf hypotheses are rejected.

\paragraph{Structural Consistency (HIR)}
A key differentiator is structural coherence. As defined by construction, P-CP achieves a HIR of exactly $0.0$ for all $\alpha$. Conversely, L-CP suffers from frequent violations of taxonomic logic, with HIR values peaking around $37\%$ at $\alpha=0.25$ (\Cref{fig:hir}). The shape of the HIR curve for L-CP is notably non-monotonic, exhibiting a ``wave'' pattern:
\begin{itemize}
	\item At very low $\alpha$ (e.g., $\approx 0.01$), sets are large. Violations are primarily ``sterile'' parent errors, where a parent set is massive (to ensure coverage) but the finer-grained \texttt{major/minor (leaf)} models manage to prune all its specific children.
	\item As $\alpha$ increases to $\approx 0.10$, sets tighten around the true class, and ``noise'' violations decrease, leading to a local minimum in HIR.
	\item As $\alpha$ increases further (towards $0.30$), the rate of empty sets rises. Crucially, emptiness is not synchronised across levels: a child model may return an empty set while the parent returns a singleton, or vice versa, creating new ``sterile'' or ``orphan'' violations. This disagreement on hard/uncertain samples drives the HIR up again.
	\item Finally, at high $\alpha$ ($>0.40$), classifiers at all levels frequently output empty sets. Since an empty parent and empty child are consistent, the violation rate drops.
\end{itemize}
This analysis highlights that while L-CP is efficient locally, it frequently presents the operator with impossible logical paths (e.g., predicting a specific \texttt{Android} version while simultaneously predicting the \texttt{OS family} as \texttt{Windows}).

\paragraph{Abstention and Usability}
\Cref{fig:empty_rate} and \Cref{fig:singleton_rate} present the empty set and singleton rates, respectively. The empty set rate remains negligible for small $\alpha$ but increases with error tolerance, reflecting the models' tendency to abstain on uncertain samples. P-CP generally produces fewer empty sets at coarser levels than L-CP (e.g., at $\alpha=0.3$, $\approx 18\%$ vs. $\approx 30\%$ for \texttt{family}), as it only discards a family if all its leaves are rejected. Regarding usability, the singleton rate rises sharply at low $\alpha$, indicating that prediction sets quickly converge to single candidates. L-CP achieves high singleton rates at the \texttt{family} level faster (lower $\alpha$) than P-CP, but both methods perform similarly at the \texttt{leaf} level, suggesting that the choice of strategy impacts usability primarily at coarser granularities.

\subsection{Discussion}
\label{subsec:discussion}

The experimental results illuminate a fundamental trade-off in hierarchical uncertainty quantification between level-wise efficiency and structural consistency.

\paragraph{Trade-off: Efficiency vs. Consistency}
L-CP optimises for the former: by calibrating each level independently, it provides the tightest possible sets that satisfy the coverage guarantee for that specific granularity. This is advantageous in scenarios where the operator is interested in a specific level (e.g., ``Is this device iOS?'') and can tolerate inconsistencies. However, the high HIR indicates that L-CP often fails to provide a coherent narrative (e.g., predicting \texttt{Android 10} while ruling out the \texttt{Android} family), which can undermine trust in automated systems. P-CP prioritises structural consistency. By enforcing nested sets, it guarantees that every prediction corresponds to a valid path in the OS taxonomy. This comes at the cost of ``inherited uncertainty'': the ambiguity at the \texttt{leaf} level propagates upwards, inflating the sets at the \texttt{family} and \texttt{major} levels.

\paragraph{Inflation Mechanism}
The observed over-coverage of P-CP at the \texttt{family} level is a direct consequence of the upward projection. In a standard classifier, a \texttt{family} might be ruled out because its probability mass is low. In P-CP, if even a single rare \texttt{minor} version of that \texttt{family} has a non-conformity score below the threshold (perhaps due to noise or calibration conservativeness), the entire \texttt{family} is retained. This ``union of ancestors'' logic effectively accumulates the false positive possibilities of the leaves, resulting in conservative sets. While this reduces efficiency, it ensures a crucial safety property: the true label is never pruned at a coarse level if it remains a plausible hypothesis at a fine level.

\paragraph{Operational Guidelines}
The choice between these methods should be dictated by the downstream application:
\begin{itemize}
	\item Automated Policy Enforcement: We recommend P-CP. Security policies often rely on hierarchical logic (e.g., ``Block all Android versions except 11+''). A consistent set allows for unambiguous logical checks. If the set contains \texttt{\{\{Android 10, Android 11\}\}}, the policy can trigger a partial block or further inspection. An inconsistent L-CP output like \texttt{\{\{family: Windows, major: Android 11\}\}} is unactionable for automated logic.
	\item Forensic Triage: L-CP may be preferable for human analysts. An analyst might care primarily about the \texttt{OS family} to route the ticket to the right team. Even if the model is confused about the \texttt{minor} version, a tight, calibrated \texttt{family} prediction is more valuable than a set inflated by \texttt{leaf} level noise. The analyst can use their domain knowledge to resolve the structural inconsistencies that L-CP produces.
\end{itemize}

\section{Conclusion}
\label{sec:conclusion}

This paper addressed the limitations of conventional Operating System fingerprinting by framing it as a hierarchical classification problem and applying Conformal Prediction (CP) to provide rigorous uncertainty quantification. We introduced and evaluated two distinct strategies: \emph{level-wise CP (L-CP)} and \emph{projection-based CP (P-CP)}, using a dataset of real-world network traffic.

Our results demonstrate that CP successfully provides valid coverage guarantees for OS identification, allowing operators to explicitly trade off prediction set size for reliability. The comparative analysis revealed a fundamental tension between local efficiency and structural consistency. L-CP yields tighter prediction sets at individual levels, making it suitable for human-in-the-loop scenarios where analysts can interpret granular outputs and resolve potential inconsistencies. Conversely, P-CP guarantees hierarchically consistent (nested) sets by construction, rendering it the superior choice for automated policy enforcement where logical contradictions must be avoided, albeit at the cost of larger prediction sets at coarser levels.

Future work will focus on mitigating the efficiency loss in consistent methods, potentially through hybrid calibration strategies or hierarchical loss functions that encourage coherence in the base models. Additionally, we plan to explore the application of Mondrian CP~\cite{vovk_mondrian_2003}---a method that ensures validity within conditional categories (e.g., per-class coverage) rather than just marginally---to better handle extreme class imbalance and to investigate adaptive recalibration mechanisms to cope with concept drift in dynamic network environments. By moving beyond point predictions to calibrated hierarchical sets, this work provides a foundational step towards more trustworthy and operationally resilient network visibility.

	\creditauthorship

	\declarationcompetinginterest

	\funding

\dataavailability

\section*{Declaration of generative AI and AI-assisted technologies in the manuscript preparation process}
During the preparation of this work, the author(s) used ChatGPT (OpenAI) and Gemini (Google) in order to support conceptualization, review drafts, and assist with writing-related tasks such as grammar, spelling, and style corrections; as well as Cursor (Anysphere) to generate and refine code snippets. After using these tools, the author(s) reviewed, validated, and edited all content as needed, and take(s) full responsibility for the content of the published article.

\end{document}